\documentclass[pra,twocolumn,a4paper,showpacs,superscriptaddress]{revtex4}

 \usepackage{latexsym}
 \usepackage{amsmath}
 \usepackage{amsfonts}
 \usepackage{graphicx}
 \usepackage{amssymb}
 \usepackage{dsfont}
 \usepackage{subfigure}

 \newcommand{\ket}[1]{\ensuremath{|#1\rangle}}
 \newcommand{\bra}[1]{\ensuremath{\langle #1 |}}
 
 \newcommand{\mc}[1]{\ensuremath{\mathcal{#1}}}
 \newcommand{\bc}{\begin{center}}
 \newcommand{\ec}{\end{center}}
 \newcommand{\mf}[1]{\boldsymbol{#1}}
 
 \newcommand{\ii}{\imath}
 \newcommand{\DP}{\Delta_p}
 \newcommand{\DC}{\Delta_c}
 \newcommand{\DR}{\delta_{R}}

\begin{document}

\title{Non-diffracting Optical Beams in a Three-level Raman System}

\author{Tarak N. \surname{Dey}}
\email{tarak.dey@gmail.com}
\affiliation{Max-Planck-Institut f\"ur Kernphysik, Saupfercheckweg 1, 69117 Heidelberg, Germany}
\affiliation{Department of Physics, Indian Institute of Technology Guwahati, Guwahati- 781 039, Assam, India}

\author{J\"org \surname{Evers}}
\email{joerg.evers@mpi-hd.mpg.de}
\affiliation{Max-Planck-Institut f\"ur Kernphysik, Saupfercheckweg 1, 69117 Heidelberg, Germany}

\date{\today}

\pacs{42.50.Gy, 32.80.Qk, 42.65.-k}

\begin{abstract}
Diffractionless propagation of optical beams through atomic vapors is investigated. The atoms in the vapor are operated in a three-level Raman configuration. A suitably chosen control beam couples to one of the transitions, and thereby creates a spatially varying index of refraction modulation in the  warm atomic vapor for a probe beam which couples to the other transition in the atoms. We show that a Laguerre-Gaussian control beam allows to propagate single Gaussian probe field modes as well as multi-Gaussian modes and non-Gaussian modes over macroscopic distances without diffraction. This opens perspectives for the propagation of arbitrary images through warm atomic vapors.
\end{abstract}

\maketitle

\section{INTRODUCTION \label{intro}}
A long sought-after goal is the all-optical processing of information, replacing state-of-the-art electronic handling.  It is well known, however, that major obstacles arise in particular at spatial scales small compared to the involved wavelengths. Light cannot be focused to arbitrarily small areas, and it cannot directly be used to image or write structures below a scale given by the involved wavelength~\cite{bornwolf,saleh}. A key origin for these restrictions is diffraction, leading, e.g., to the Rayleigh limit for the case of imaging~\cite{rayleigh}. Due to the enormous significance for applications, e.g., in nano- or life-sciences, a number of schemes have been invented to circumvent or even surpass the standard limits in a various setups. Examples are sub-wavelength microscopy~\cite{hell,hettich,scully,distance,distance2,gulfam,complicated,rabi,kapale,Herkommer,Walls,Rempe}, and optical sub-wavelength  lithography~\cite{prop1,prop2a,prop2b,prop2c,prop2d,prop3,rabi-litho}. But more fundamentally, diffraction also already impacts the propagation of a light beam in free space.  For example, a Gaussian probe beam with a certain width at the focal plane  spreads as it propagates away from the focal plane, on a length scale given by the Rayleigh length~\cite{saleh}. This paraxial diffraction spreading occurs in free space as well as inside a medium, and affects applications such as the transmission of small images, the transfer of small mask structures to the target surface in lithography, or the steering or optical manipulation of light beams.

It was recognized that this spreading due to paraxial diffraction can be decreased, increased or completely eliminated by propagating the light beam through a medium with spatially varying optical properties. This can be achieved via the use of external coherent control fields, inducing a spatially varying susceptibility and inducing phase shifts via different physical processes~\cite{prop4}. The spatial profile of a strong control beam renders the medium inhomogeneous along the transverse direction. Since absorption and dispersion of the medium are crucially dependent on the intensity and the shape of the strong control beam, the control beam will strongly affect the propagation of the probe beam. The ability to control the spatially varying susceptibility of the medium via external fields holds the key to eventually being able to control the spreading of optical beams. These and related techniques are of significance also because they have the potential of complementing optical information transmission with all-optical information processing, thereby replacing the currently ubiquitous electronic information processing.

\begin{figure}[b]
\includegraphics[width=6.0cm]{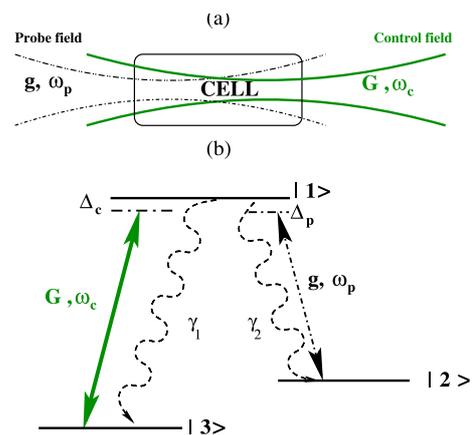}
\caption{\label{Fig1}
(Color online) Diffractionless beam propagation through a vapor cell. (a) shows a schematic setup to study the propagation of a probe laser field through an atomic vapor cell. The propagation dynamics is controlled using an additional control laser field with a Laguerre-Gaussian transversal profile. (b) Three-level $\Lambda$ configuration of the vapor atoms.}
\end{figure}

Thus it is not surprising that a great number of schemes both in theory and in experiment demonstrated the ability to optically control the propagation of light.  Most of these works can be understood in analogy to a waveguide, confining the light in specific modes transversally.  This could enable one to implement applications originally proposed for physical waveguides~\cite{wg1,wg2} with all-optically induced waveguides. For example, it was shown that spatial variations in a medium induced by a control laser field can lead to focusing or de-focusing of a probe beam~\cite{focusing1, focusing2,focusing3, focusing4}. An optically written waveguide realized by applying a Laguerre-Gaussian control field near-resonantly to an atomic medium to propagate a probe beam was demonstrated in~\cite{prop5} and discussed in~\cite{theory1, theory2}. Waveguiding in a warm rubidium vapor based on an off-resonant Raman configuration was reported in~\cite{prop6}.  A signal beam was shown to propagate with small spot size over several diffraction lengths. In a variant of such setups, the spatial profile of a control beam can be cast onto a probe beam. Recently, this concept was implemented in Rubidium atomic vapor, based on coherent population trapping~\cite{prop7}. In this experiment, a spatial profile was transferred from the coupling beam onto the probe beam, which was found to exhibit feature sized five times smaller than the diffraction-limited size.

However, all of these schemes suffer from the fact that the modes other than the specific supported mode cannot propagate without diffraction, or even not at all. This prohibits the desired transmission of arbitrary images. In order to overcome this limitation, very recently, a scheme for the direct propagation of arbitrary probe beam images without diffraction was put forward theoretically~\cite{prop8} and verified experimentally~\cite{prop9}. This setup is based on an atomic three-level system reminiscent of electromagnetically induced transparency, but assumes an explicit breaking of the usual two-photon resonance condition such that strong absorption occurs. The underlying physics in this setup is fundamentally different from previous setups, as it relies on the movement of the atoms in order to compensate the diffraction, rather than on a stationary waveguide.
A method to propagate optical images with reduced diffraction based on the saturated absorption mechanism in a two-level atom was proposed in~\cite{dey1}.
Future ideas along this line therefore can be expected to advance key applications in optical information transmission and processing, as well in reading and writing at the sub-wavelength scale.

In this paper, we discuss the light propagation through waveguides written by a spatially structured control laser field into a warm atomic vapor. The Laguerre-Gaussian control field and the probe laser field couple with the atoms to form a three-level system in $\Lambda$ configuration as shown in Fig.~\ref{Fig1}. We start by deriving analytical results for the steady state susceptibility experienced by the probe beam. We then consider a thermal medium by averaging the susceptibility over a Maxwellian velocity distribution of the vapor atoms. The probe beam propagation dynamics is analyzed by numerically solving the Maxwell-Schrödinger equation including diffraction and dispersion in two spatial dimensions transversal to the propagation direction, the spatial profile of the control field, and the thermal motion of the vapor atoms. We find that for suitable parameter choices, a single Gaussian probe field mode can be propagated diffraction-less or even with sub-diffraction through the warm vapor. We then proceed to analyze more complex mode structures of the probe beam, and show that a Gaussian double pulse can be propagated without distortion or diffraction through the optically written waveguide. Finally, we demonstrate that a beam formed out of multiple secant-hyperbolic shaped pulses can be propagated diffraction-less through the warm atomic vapor. The results on non-Gaussian beam propagation open perspectives for the propagation of arbitrary images in warm atomic vapors.

The paper is organized as follows. In Sec.~\ref{theory}, we introduce our model, discuss the susceptibility of a thermal vapor, and describe the Maxwell-Schr\"odinger equation used to analyze the propagation dynamics of the probe beam through the optically written waveguide. In Sec.~\ref{tailor}, we discuss the static properties of the medium susceptibility under the action of the spatially structured control laser beam. In Sec.~\ref{prop}, we then discuss our results on the propagation of different probe beams through the optically written waveguide. Finally, Sec.~\ref{summary} discusses and summarizes our results.

%
%
\section{\label{theory}Theoretical considerations}
\subsection{The model}
We consider the geometry as shown in Fig. \ref{Fig1}. A weak probe laser field with frequency $\omega_p$ is focused at the entry of an atomic vapor, and co-propagating with an additional control laser field with frequency $\omega_c$ focused at the back of the vapor cell. The cell is filled with a gas of three level atoms in $\Lambda$-configuration as shown in Fig. \ref{Fig1}(b). The electric fields of the two beams can be defined as
\begin{subequations}
\label{probe}
\begin{align}
\mf{E}_j(\mf{r},t) & = \mf{E}_j^{(+)}(\mf{r},t) + \text{c.c.}\,, \\
 \mf{E}_j^{(+)}(\mf{r},t)&= \mf{e}_{j}\mc{E}_{j}(\mf{r},t)~e^{-i\omega_j t + i \mf{k}_j\cdot\mf{r}} \,,
\label{probePlus}
\end{align}
\end{subequations}
where $\mc{E}_j$  is the slowly varying envelope and $\hat{e}_j$ is unit polarization vector of the field. The index $j\in \{c,p\}$ denotes the probe $(p)$ or control $(c)$ field, respectively. The polarization vectors of the beams are chosen in such a way that the control beam $\mf{E}_c$ is tuned to the atomic transition $\ket{1}\leftrightarrow\ket{3}$ while the probe beam $\mf{E}_p$ is coupled to the atomic transition $\ket{1}\leftrightarrow\ket{2}$. The Hamiltonian of the $\Lambda$-system under electric-dipole and rotating-wave approximation can then be written as
\begin{subequations}
\label{Hschroed}
\begin{align}
H =& H_0 + H_I\,,\\
H_0 =& \hbar\omega_{13}\ket{1}\bra{1} + \hbar\omega_{23} \ket{2}\bra{2}\,,\\
H_I =& - ( \ket{1}\bra{2} \mf{d}_{12}\cdot\mf{E}_p^{(+)} \nonumber \\
 &  + \ket{1}\bra{3} \mf{d}_{13}\cdot\mf{E}_c^{(+)}   \,+\,\text{H.c.})\,,
 \end{align}
\end{subequations}
where $\omega_{j3}$ denotes the resonance frequency on the transition $\ket{j}\leftrightarrow\ket{3}$, $\mf{d}_{1j}=\bra{1}\mf{\hat{d}}\ket{j}$ is the  matrix element of the electric dipole moment operator $\mf{\hat{d}}$, and we have set the energy of state $\ket{3}$ to zero. We now make use of the unitary transformation
\begin{subequations}
\begin{align}
  W&=e^{-\frac{i}{\hbar}U t} \,,\\
  U&=\hbar \omega_c \ket{1}\bra{1} + \hbar (\omega_c-\omega_p)\ket{2}\bra{2}\,,
\end{align}
\end{subequations}
to rewrite the Hamiltonian as
\begin{align}
 V  = & \hbar\Delta_c\ket{1}\bra{1} + \hbar(\DC-\DP)\ket{2}\bra{2}  \notag \\
 & -\hbar \left(G\ket{1}\bra{3} + g  \,\ket{1}\bra{2}    \,+\,\text{H.c.}\right)\,.
 \label{H}
\end{align}
Here, $\Delta_j$ for $j\in \{c,p\}$ are the detunings of the fields defined as
\begin{subequations}
\begin{align}
\label{detuning}
 \Delta_p &= \omega_{12} - \omega_p\,,\\
\Delta_c &= \omega_{13}- \omega_c\,.
 \end{align}
\end{subequations}
The Rabi frequencies of the probe and control fields $2g$ and $2G$ are related to the slowly varying amplitudes of ${\cal E}_p$ and ${\cal E}_c$ as
\begin{subequations}
 \begin{align}
\label{Rabi_fre}
 2g&=\frac{2\vec{d}_{12}\cdot\vec{\mathcal{E}}_{\rm
 {p}}}{\hbar}\,,\\
 2G&=\frac{2\vec{d}_{13}\cdot\vec{\mathcal{E}}_{\rm{c}}}{\hbar}\,.
 \end{align}
\end{subequations}

As shown in Fig.~\ref{Fig1}(b), the upper atomic state $|1\rangle$ decays to $\ket{3}$ with rate $\gamma_1$, and to $\ket{2}$ with rate $\gamma_2$. Furthermore,  $\Gamma_{23}$  is the decay rate of the ground state atomic coherence. In order to incorporate these incoherent processes, we use the master equation approach to describe the dynamics of the atomic populations and coherences of the three level atoms. From Eq.~(\ref{H}), the equations of motion for the elements of the density matrix $\rho$  can be written as
\begin{subequations}
\label{Full_density}
\begin{align}
 \dot{\rho}_{11}=&-(\gamma_1+\gamma_2)\rho_{11}+ \ii g \rho_{21}
 + \ii G \rho_{31} \nonumber \\
& -\ii g^{*} \rho_{12}-\ii G^{*} \rho_{13} \,,\\
 \dot{\rho}_{22}=&  \gamma_2\rho_{11} + \ii g^* \rho_{12}  - \ii g
 \rho_{21}\,,\\
 \dot{\rho}_{12}=&-\left[\frac{1}{2}(\gamma_1 + \gamma_2) + \ii \DP\right]\rho_{12} + \ii g  \rho_{22} \nonumber\\
&+ \ii G \rho_{32} - \ii g\rho_{11} \,,\\
 \dot{\rho}_{13}=&-\left[\frac{1}{2}(\gamma_1 + \gamma_2) + \ii \DC\right]\rho_{13}
 + \ii g\rho_{23} \nonumber\\
& + \ii G\rho_{33} - \ii G\rho_{11}\,,\\
 \dot{\rho}_{23}=& -[\Gamma - \ii (\Delta_p-\Delta_c)]\rho_{23}
 + \ii g^*\rho_{13} - \ii G\rho_{21}\,.
 \end{align}
\end{subequations}
The third population follows from the conservation law
$\rho_{11}+\rho_{22}+\rho_{33}=1$.
In the following, we for simplicity restrict the analysis to the case $\gamma_1=\gamma_2=\gamma$.

\subsection{Susceptibility of a thermal medium}

In the steady state, the density matrix equations~(\ref{Full_density}) can be solved analytically. We are in particular interested in the steady state value of the atomic coherence $\rho_{12}$, which determines the probe field susceptibility $\chi$  of the medium at frequency $\omega_p$. The analytical expression for $\chi$ to all orders in the control and probe Rabi frequencies is given in Appendix~\ref{susceptibility}. The susceptibility crucially depends on the two photon detuning of the Raman system  defined as
\begin{align}
\DR=\DP-\DC\,.
\end{align}
For a warm atomic medium, the thermal motion of the atoms causes inhomogeneous broadening of the atomic spectra. This broadening plays a important role when considering the optical properties of the medium to determine the probe beam propagation dynamics through the medium \cite{kash,kochar}. We incorporate the motion in the susceptibility $\chi$ in Eq.~(\ref{mollow_chi}) by defining velocity-dependent detuning of a given atom
\begin{subequations}
\label{detuning_dop}
\begin{align}
\DP(v)&=\DP - kv\,,\\
\DC(v)&=\DC - kv\,,
\end{align}
\end{subequations}
where $v$ is the velocity of the atom. For simplicity, we have assumed assume $k\approx k_p\approx k_c$. The velocities of all atoms obey the Maxwellian distribution
\begin{equation}\label{Max_Boltz1}
{\rm P}(kv)d(kv)=\frac{1}{\sqrt{2\pi{\rm D^2}}}e^{-\frac{(kv)^2}{2{\rm
D}^2}}d(kv),
\end{equation}
with $D$ is the Doppler width defined by
\begin{equation}
D=\sqrt{\frac{K_{B}T{\omega}^2}{Mc^2}}
\end{equation}
at temperature $T$. For a warm atomic system, the averaged susceptibility  is given by
\begin{equation}\label{chi_average}
\langle\chi\rangle=\int_{-\infty}^{\infty} \chi(kv) P(kv)d(kv)\,.
\end{equation}

\subsection{Beam propagation dynamics}

In general, the medium susceptibility is a function of the position in the vapor cell due to the spatial structure of the applied laser fields. In order to account for the spatial variations, we study the propagation of the light beams through the medium using Maxwell's equations. For this, we use the slowly varying envelope approximation, and assume that the changes along the longitudinal direction $z$ of the probe envelope is very small compared to the changes which occur along the transverse directions $x,y$. The beam propagation equation for the Rabi frequency $g$ of the probe field under the paraxial wave approximation then follows as
\begin{align}
\label{max1}
 \frac{\partial g}{\partial z} = \frac{i c}{2 \omega}
\left(\frac{\partial^2}{\partial x^2}
+\frac{\partial^2}{\partial y^2}\right)g + 2 i \pi k \langle\chi\rangle g\,,
\end{align}
where $\omega=ck$ and $c$ is the speed of light in free space. The thermal motion of the atoms is included in the the beam propagation by considering the average response of the susceptibility $\langle \chi \rangle$. The terms in the parentheses in Eq.~(\ref{max1}) are responsible for the diffraction of the probe beam. In order to study the effect of diffraction and spatial dispersion on the probe beam propagation, we solve Eq.~(\ref{max1}) numerically using a higher order split step operator method \cite{Shen}.

\section{\label{results}Results}

\begin{figure}[t]
\includegraphics[width=6 cm]{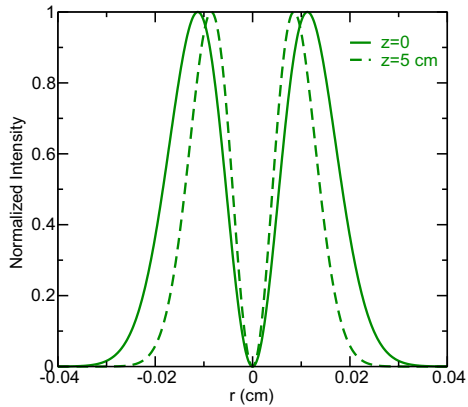}
\caption{\label{Fig2}
(Color online) Normalized spatial intensity profile of the control laser field. The solid (dashed) line shows the profile at the entry (exit) plane of the vapor cell. $r$ is the radial distance from the beam center. The beam waist and the Rayleigh length of the Laguerre-Gaussian beam are chosen as 120~$\mu$m and 5.7~cm, respectively.}
\end{figure}

\subsection{\label{tailor}Tailoring the medium susceptibility using a shaped control beam}

In order to illustrate how the spatial structure of the control field enables us to guide the weak probe beam through warm atomic vapor, we first study the behavior of the susceptibility of the medium as a function of the spatial variation of the control field intensity. We chose a doughnut shaped Laguerre-Gaussian (LG) control field to induce a waveguide-like refractive index pattern inside the atomic vapor. The Rabi frequency of the LG control field can be written as
\begin{subequations}
\label{control_field}
\begin{align}
 G & = G^{\rm 0} \frac{{w_{c} r}}{{w_z^2 }}\exp\left[ { - \frac{{ikr^2 }}{{2q}} + i\theta } \right]\,, \\
 r &= \sqrt {x^2  + y^2 }\,,\label{Beam_waist}\\
 w_{z}&=w_{c} \sqrt {1 + \left( {\frac{z-z_{0}}{{z_{R} }}} \right)^2}\,\\
q&=\imath z_{R} -z + z_0\,,\\
\theta&=\tan ^{ - 1} \left[ {\frac{y}{x}} \right]\,,\\
z_{R}  &= \frac{{\pi w_{c}^2 }}{\lambda }\,,
\end{align}
\end{subequations}
where $z_0$ and $z_{R}$ are the location of the beam waist and the Rayleigh length, respectively.

Figure~\ref{Fig2} shows the intensity variation of the control  field against radial position $r$ at the entry and exit faces of the medium. For this figure, we have considered a vapor cell of length 5~cm and filled with Rb atoms. The intensity of the control field approaches zero for $r\to 0$ and maxima occur at $|r|=0.0115$~cm at the entry face of the medium. It should be noted that there is no significant change in the structure of the control field envelope throughout the propagation through the medium. This is consistent with the results in the experiment by Howell et al~\cite{prop6}. From the point of view of an additional probe beam, the control field structure renders the susceptibility inhomogeneous along the transverse direction.

\begin{figure}[t]
\centering
   \includegraphics[width=0.9\columnwidth] {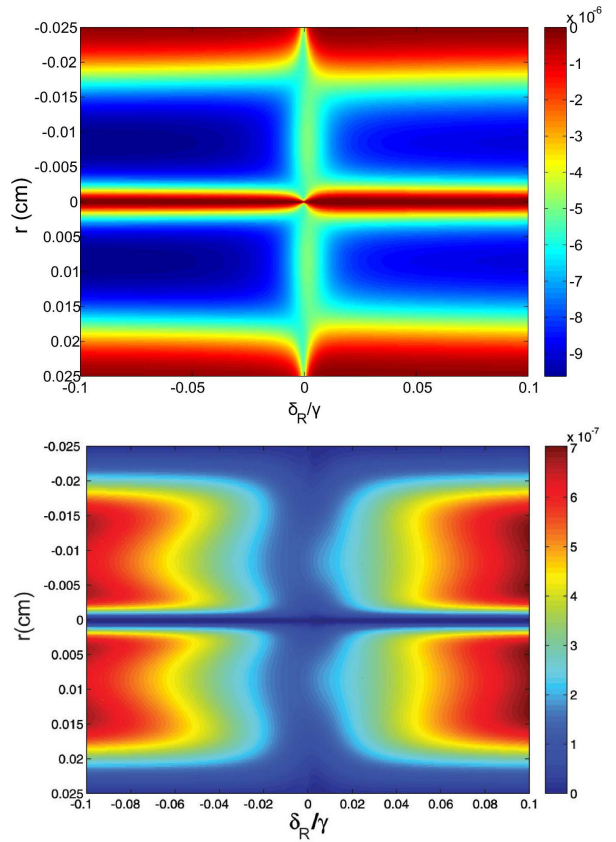}
\caption{\label{fig:Fig3}(Color online) Real part (a) and imaginary part (b) of the averaged susceptibility $\langle\chi\rangle$ in a thermal vapor obtained from Eq.~(\ref{chi_average}) in the presence of the LG control beam. The results are shows as a function of the radial position $r$ in the beam and the Raman detuning $\delta_{R}/\gamma$. The other parameters are chosen as $G^0=1\gamma$, $w_{c}=120\mu$m, $g^0=0.2\gamma$, $\Delta_p=-170\gamma$, $D=70\gamma$, $N=10^{12}$ atoms/cm$^{3}$ and $\Gamma=0.001\gamma$.}
\end{figure}

\begin{figure}[t!]
\includegraphics[width=6.0cm]{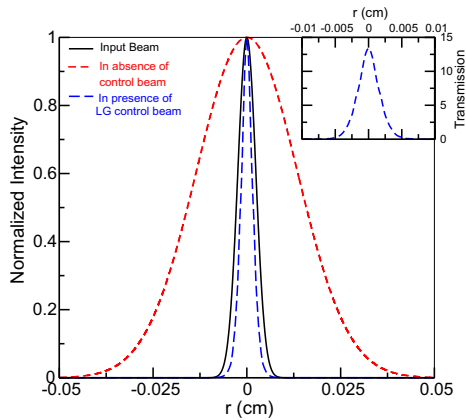}
\caption{\label{Fig4}
(Color online)
Normalized intensity of the transmitted probe beam plotted against the radial distance from the beam center $r$. The solid black line shows the profile at the input of the vapor cell. The other two curves show the profile at the output after a propagation length of 5~cm. The red short-dashed curve is obtained without control field, whereas the blue long-dashed line is obtained with a LG control laser field. The inset shows the variation of the unnormalized output intensity of the probe beam. The initial width and Raman detuning of the off-resonant probe field are chosen as $w_{p}$=48$\mu$m and $\delta_{R}=-0.015\gamma$, respectively. All other parameters are as in Fig.\ref{fig:Fig3}.}
\end{figure}

\begin{figure}[t]
\centering
\subfigure[]
 {
   \includegraphics[width=0.9\columnwidth]{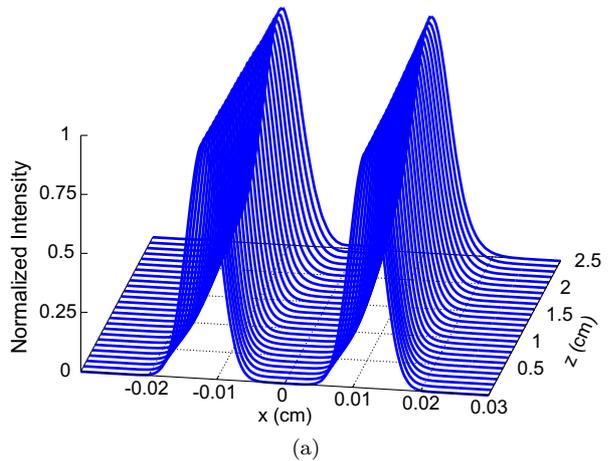}
   \label{fig:Fig5a}
 }
\subfigure[]
 {
 \includegraphics[width=0.9\columnwidth]{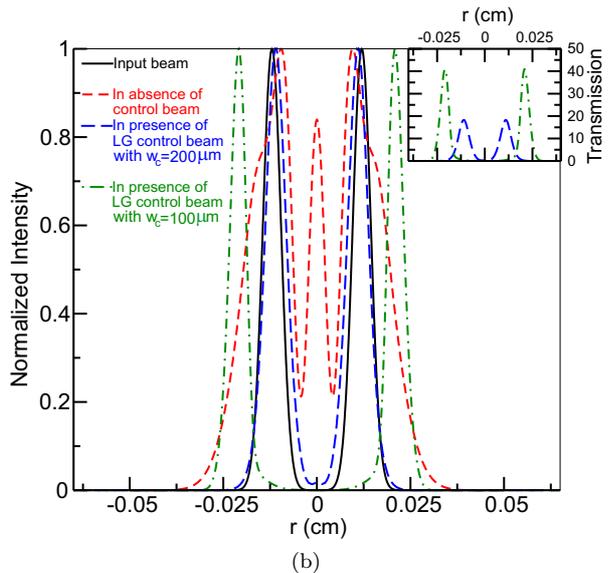}
   \label{fig:Fig5b}
 }
\caption{\label{Fig5}(Color online) Normalized intensity of the transmitted probe beam. In (a), the intensity profile is shown against the transverse coordinate $x$ in the $y=0$ plane at different propagation distances $z$. (b) compares the probe beam profile with and without the LG control field at the output of the vapor cell after a propagation of 2.5 cm. The variation of the unnormalized output intensity of the probe beam is shown in the inset. The initial amplitude and width of the off-resonant control and probe fields are chosen as $G^{0}=0.75\gamma$, $g^{0}=0.2\gamma$, $w_{p}=48\mu$m, $w_{c}=100\mu$m and $200\mu$m. All other parameters are the same as in Fig~\ref{fig:Fig3}.}
\end{figure}

\begin{figure}[t]
\centering
\subfigure[]
 {
   \includegraphics[width=0.9\columnwidth]{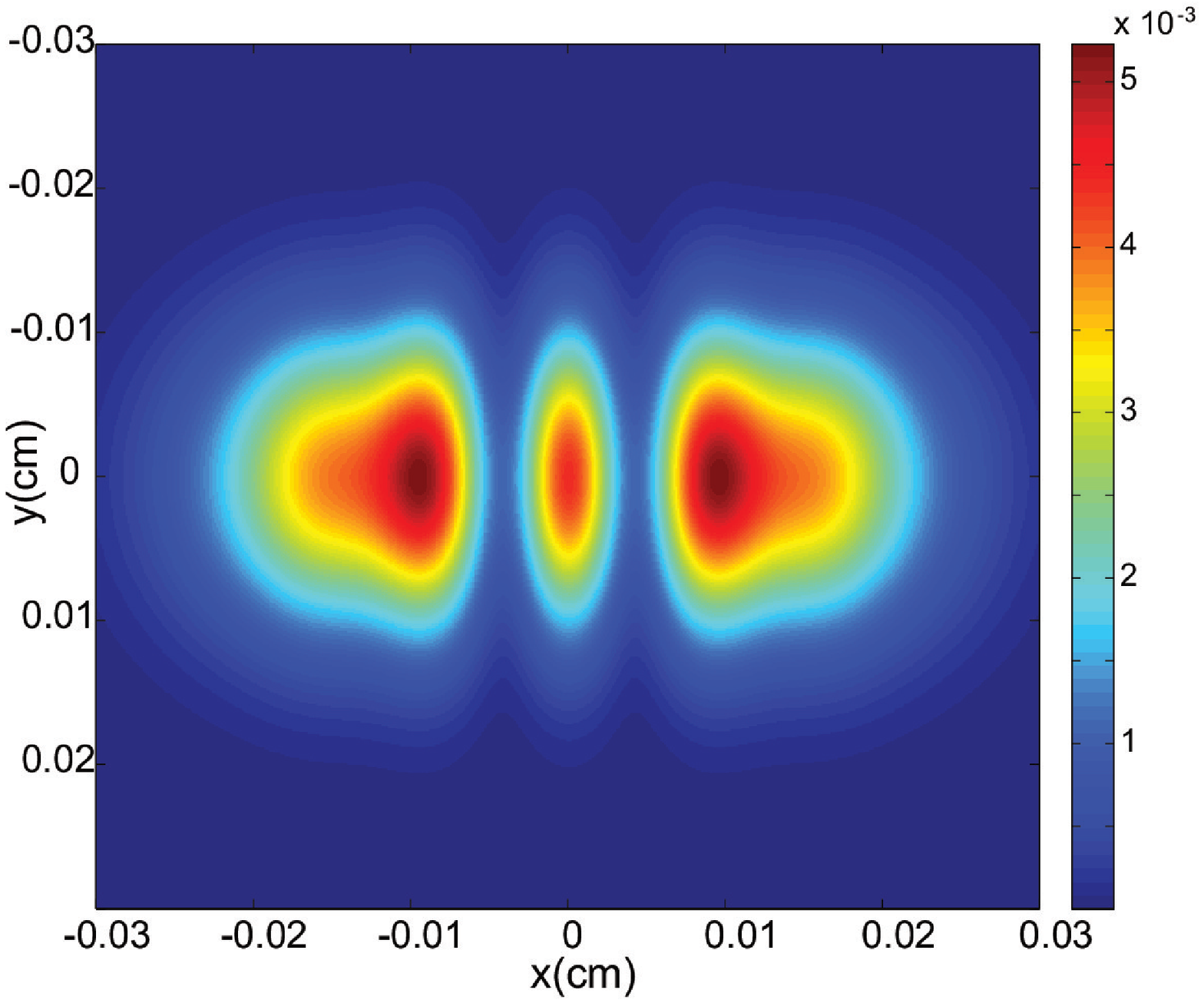}
   \label{fig:Fig6a}
 }
\subfigure[]
 {
   \includegraphics[width=0.9\columnwidth]{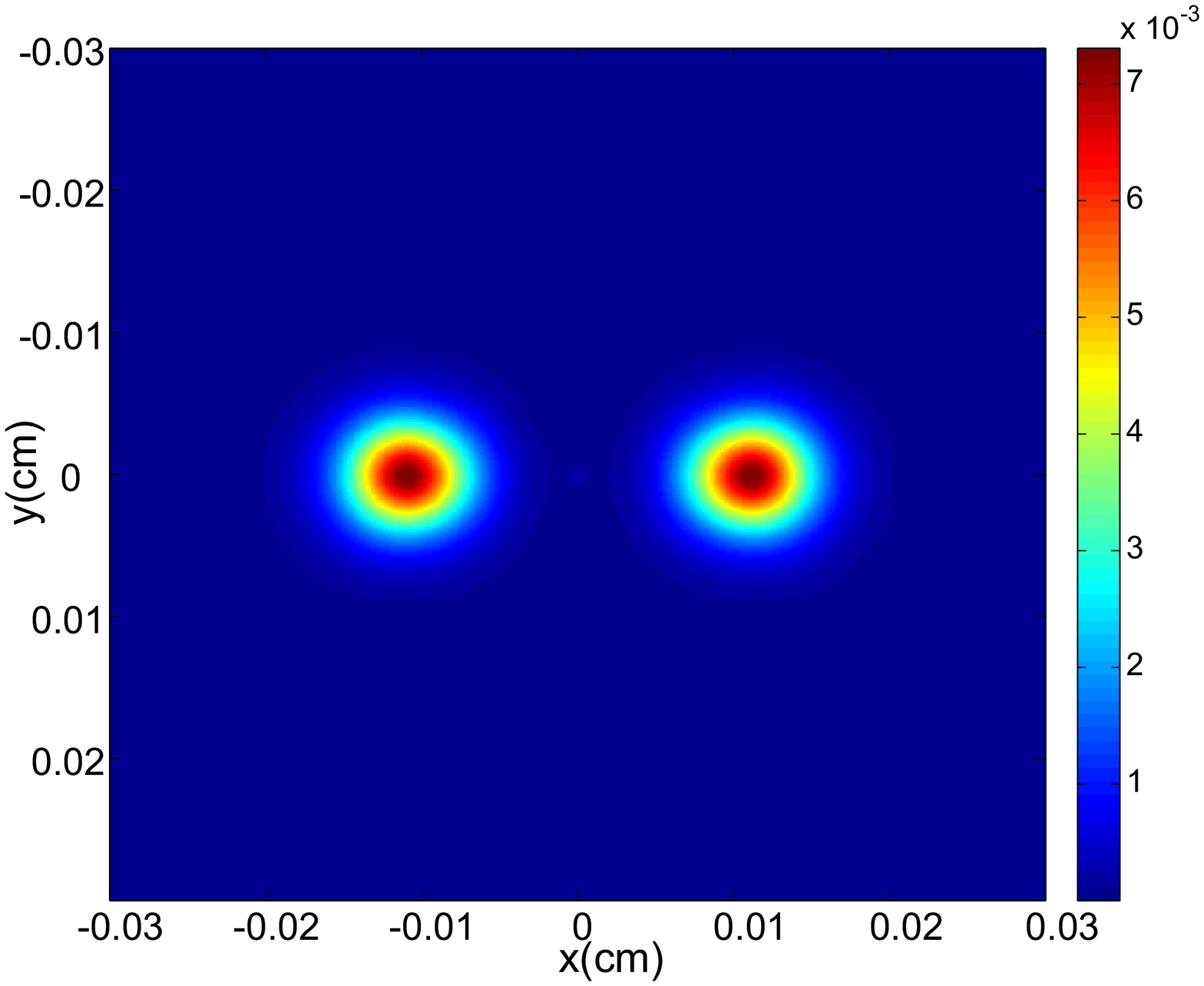}
   \label{fig:Fig6b}
 }

\caption{\label{Fig6}(Color online) Intensity of the transmitted probe beam in the $x-y$ plane transversal to the beam propagation direction at the output of the vapor cell in the absence (upper panel) and presence (lower panel) of the LG control beam, respectively. The parameters are as in Fig.~\ref{Fig5}.}
\end{figure}

Next, we have solved Eq.~(\ref{chi_average}) numerically in the presence of a continuous wave probe field to analyze the spatial inhomogeneity of the susceptibility. Figure~\ref{fig:Fig3}  shows surface plots of the real and imaginary parts of $\langle\chi\rangle$ as a function of the radial position $(r)$ and the Raman detuning $(\delta_R)$. For the calculation, we have fixed the single photon probe detuning to $\Delta_p=-170\gamma$, and use typical parameters for the $^{85}$Rb transition: $\gamma=3\pi\times10^6$ rad/sec, $D=1.33\times10^9$ rad/sec, and $N=10^{12}$ atoms/cm$^3$. It can be seen from Fig.~\ref{fig:Fig3}(a) that the refractive index is high at radial positions symmetrically located around $r=0$ at which the control intensity is high. Towards the beam center and for larger radial distances $|r|$, it decreases. The difference between the maximum and the minimum of the refractive index is of order 10$^{-5}$, which is in agreement with the experimental results of Howell et. al~\cite{prop6}. This shows that a LG control beam allows one to mimic a fibre like refractive index profile inside the atomic medium. From Fig.~\ref{fig:Fig3}(b) we find that the absorption of the medium has a minimum at $\delta_{R}=-0.02\gamma$, whereas an increase or decrease of the Raman detuning from this value results in increased absorption by the medium.

\subsection{\label{prop}Probe beam propagation in the tailored medium}

The analytical expressions Eqs.~(\ref{control_field}) for the control field Rabi frequency allow us to study the effect of  the spatial structure of the control beam on the probe beam propagation dynamics. For this, we have included the control field profile Eqs.~(\ref{control_field}) in the propagation equation~(\ref{max1}).

\subsubsection{Single Gaussian probe field mode}
First, we study the propagation dynamics of a Gaussian probe beam. The shape of the probe beam is defined as
\begin{subequations}
\begin{align}
g &= g^0 e^{-r^2/w_{1}(z)^2}\,,\\
w_{1}(z)&=w_{p}\sqrt{1+\left(\frac{z}{z_R}\right)^2}\,,
\end{align}
\end{subequations}
where $w_{1}(z)$ is the beam width.
Numerical results are shown in Fig.~\ref{Fig4}, for a Rayleigh length of the probe beam chosen as  $0.9$~cm. The figure compares the propagation of the probe beam through the medium in the presence and the absence of the control field. In free space or in the absence of the control beam, we observe from Fig.~\ref{Fig4} that the probe beam spreads to nearly 5.64$w_{p}$ during the propagation over a distance of 5~cm. The most intuitive way to understand the spreading of the beam is by decomposing it into a set of plane waves. In the course of the propagation, each plane wave acquires its own unique phase shift. The superposition of the phase shifted plane waves at the output gives rise to spreading. In contrast, for Bessel, Mathieu and Airy beams each constituent plane wave acquires exactly the same phase shift. In these cases, diffractionless propagation is possible through free space.

Analogously, the manipulation of diffraction-induced spreading is possible by changing the spatial dispersion of the medium. This spatial dispersion can be substantially modified by employing a suitable structure of the control field. In particular, a suitable structure of the control field can cancel or even reverse the effect of diffraction. Our numerical results in Fig.~\ref{Fig4} obtained with the LG control field show that the width of the probe beam is 37$\mu$m after a 5~cm propagation distance. Thus, the width of the output probe beam in the presence of the LG control beam is narrower than the width of the input probe beam, demonstrating that the LG control field can be used to reverse and even over-compensate the effect of diffraction. These results are in good agreement with the experiment of Howell et al~\cite{prop6}. We also notice from the inset of Fig.~\ref{Fig4} that the integrated transmission intensity of the output probe beam is 44$\%$ of the integrated intensity of the input probe beam. This indicates that the probe beam can be guided efficiently in an atomic vapor.

\subsubsection{Double Gaussian probe field mode}
Next, we study the propagation of a beam with two Gaussian modes, using the same method as for the single Gaussian probe beam. Figure~\ref{fig:Fig5a} shows the normalized intensity of both transmitted Gaussian beams in the $y=0$ plane along the $z$ axis for a propagation length up to $2.5$~cm in the presence of the LG control field. In comparison, without the control beam or in free space the incident probe beam undergoes spreading due to its geometrical shape as shown in Fig.~\ref{fig:Fig5b}. It can be seen from Fig~\ref{fig:Fig5b} that again, the spreading due to diffraction can be eliminated using the LG control beam. The inset of Fig.~\ref{fig:Fig5b} shows the unnormalized transmission of the two Gaussian beams in the $y=0$ plane. It is clear from the inset of the
Fig.~\ref{fig:Fig5b} that the transmitted energy  of probe beam gets reduced by $~50\%$ while
the width of the LG control beam is changed from $w_{c}=100~\mu$m  to $200~\mu$m.
We also notice that the spacing between two peaks of the output probe beam can be increased by decreasing the width of the LG control beam. This shift in the probe beam peak positions arises if the probe beam spectrum is located on the sides of the control field intensity pattern, rather than symmetrically around the control field intensity maximum. As an application, the finesse of the output probe beam can be controlled by changing the width of the LG control beam.

\begin{figure}[t]
\centering
\subfigure[]
 {
   \includegraphics[width=0.9\columnwidth]{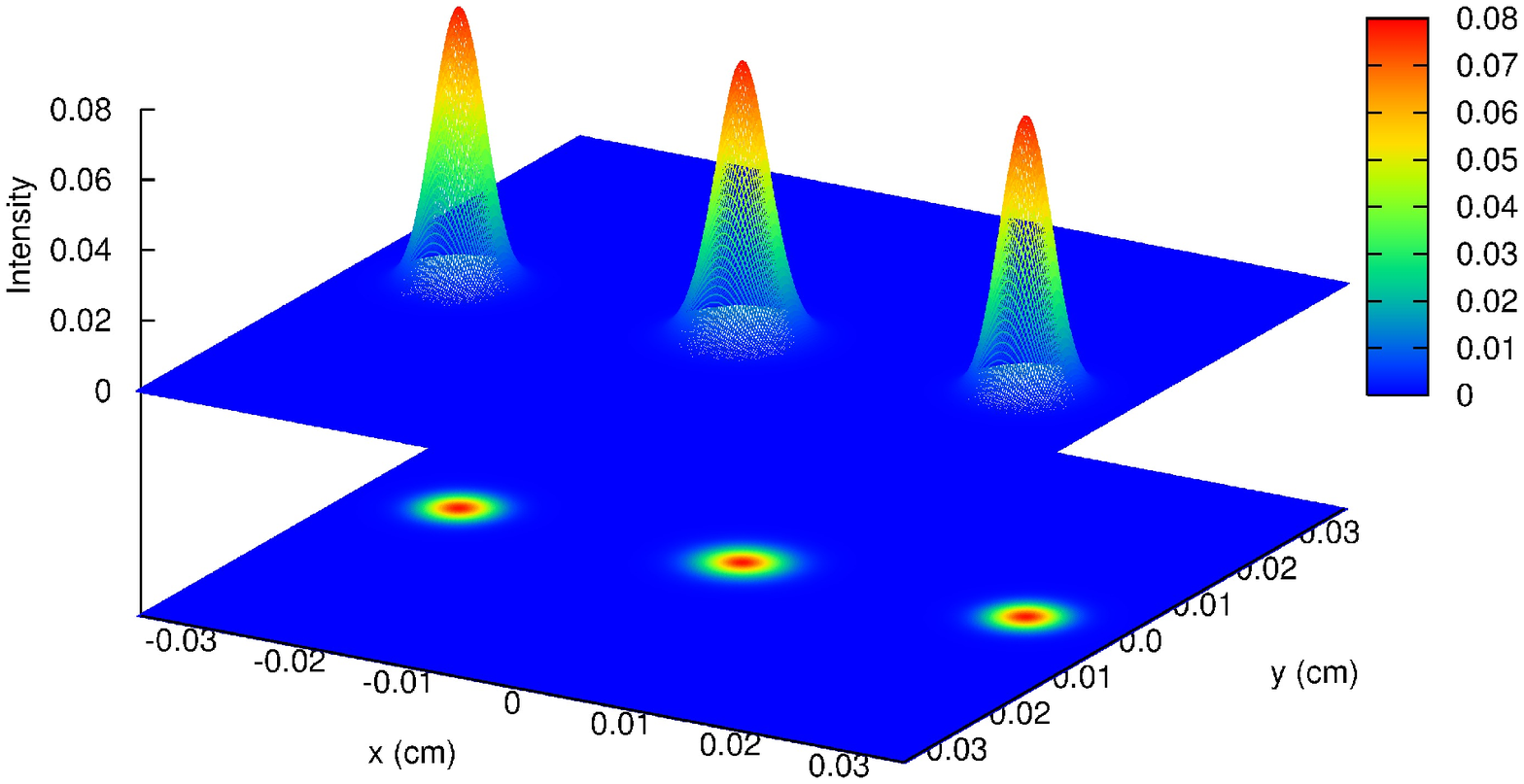}
   \label{fig:Fig7a}
 }
\subfigure[]
 {
   \includegraphics[width=0.9\columnwidth]{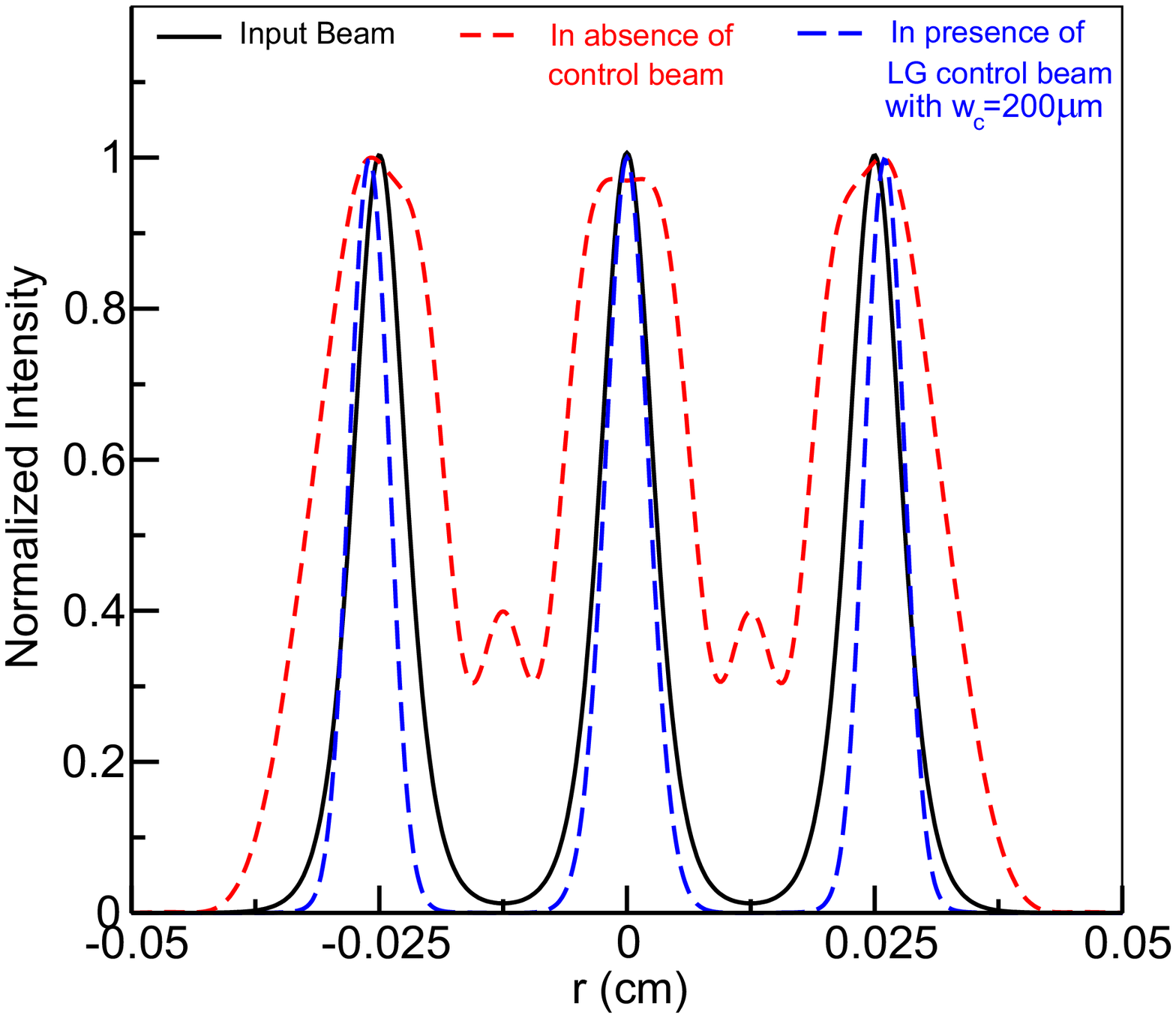}
   \label{fig:Fig7b}
 }

\caption{\label{Fig7}(Color online) (a) Image of a non-Gaussian multi-peak beam  at the output of a vapor cell with length L=2.5 cm. (b) shows a comparison of the cases with and without control beam at the output of the vapor cell for the non-Gaussian beam. The parameters are as in Fig.~\ref{Fig5} expect that the non-Gaussian probe beam has width 35 $\mu$m and the LG control beam has width 200 $\mu$m. }
\end{figure}

Figure~\ref{Fig6} shows the radial distribution of the output probe beams at $z=2.5$~cm, for a medium without and with the LG control beam, respectively. As in Fig.~\ref{fig:Fig5b} it can be seen that in free space or without the control beam, the beam profiles are severely distorted. In contrast, with the LG control beam, the shape of the two Gaussian beams is preserved. This shows that the spatial structure of the control field can be used to transmit more complicated structures over a large distance without much diffraction and attenuation.

\subsubsection{Non Gaussian probe field mode}
We proceed with a probe beam formed out of multiple secant-hyperbolically shaped patterns as an example for the transmission of a non-Gaussian field mode. The probe beam is defined as
\begin{align}
\label{NonGaussian}
g = g^0 \sum\limits_{i=0}^n \textrm{sech}\left[\frac{\sqrt{(x-x_i)^2+y^2}}{w_{1}}\right],
\end{align}
where $n$, $x_i$ and $w_i$ are the total number, the center, and the widths of the peaks, respectively. For simplicity, in our calculation, we assume that the individual peaks have equal widths $w_i$. Figure~\ref{fig:Fig7a} shows the radial variation of probe beam intensity in the presence of the LG control beam at the exit plane of the vapor at $z=2.5$~cm. A corresponding comparison of the non-Gaussian beam propagation through the medium with and without the control beam is shown in Fig.~\ref{fig:Fig7b}. It is important to note from  Fig.~\ref{fig:Fig7b} that the output beam retains sub-peaks of secant-hyperbolic profile as they were present in the input beam. Thus the LG control beam allows to propagate Gaussian as well as non-Gausssian beams through the medium without diffraction.

\section{\label{summary}Summary}
We have discussed diffraction-less propagation of Gaussian and non-Gaussian probe beams through a warm atomic vapor suitably tailored using an additional control beam. The Laguerre-Gaussian control beam writes a spatially modulated index of refraction in the vapor, which leads to a modified dispersion experienced by the probe beam, and subsequently to diffraction-less propagation of the probe beam. We have shown that probe beams of single Gaussian, multiple Gaussion, and multiple non-Gaussian structure can be propagated through the medium without diffraction. 

\acknowledgments
TND thanks P. K. Vudyasetu for correspondence on off-resonant absorption in a $\Lambda$-type system.

\appendix

\section{\label{susceptibility}Steady state susceptibility}
The steady-state susceptibility is given by
\begin{align}
\label{mollow_chi}
\chi=\frac{\textrm{N}|d|^2}{\hbar} \frac{\mathcal N}{\mathcal D},
\end{align}
where the numerator $\mathcal N$ and the denominator $\mathcal D$ evaluate to
\begin{align}
\mathcal N =&
|G|^2\left[\gamma
 \left(\ii \Gamma-\DR\right)\left\{|G|^2+(\gamma-\ii \DP)
 (\Gamma-\ii \DR)\right\} \right.\nonumber\\
 &\left. +|g|^2\left\{\gamma(\ii\Gamma-\DR)+\Gamma(\DC+\DP)\right\}\right]\,, \\
D=&\gamma |g|^6 + |g|^4\left\{3|G|^2(\gamma+2\Gamma)+2\gamma\left(\gamma\Gamma+\DR\DC\right)\right\} \nonumber\\
& + \gamma |G|^2\left\{(\Gamma^2+\DR^2)(\gamma^2+\DP^2)+2|G|^2(\gamma\Gamma+\DR\DP) \right . \nonumber \\
&\left . +|G|^4\right\} + |g|^2 \left\{3 |G|^4 (\gamma +2 \Gamma )\right. \nonumber \\
&\left . + \gamma  \left(\gamma ^2+\DC ^2\right)
 \left(\Gamma ^2+\DR^2\right)+\left(4 \gamma ^2 \Gamma +\gamma
 \left(6 \Gamma ^2+4\DR^2\right) \right. \right. \nonumber \\
& \left. \left. +\Gamma  (\DC +\DP )^2\right) |G|^2 \right\}\,.
\end{align}


\begin{thebibliography}{99}

\bibitem{bornwolf}M. Born and E. Wolf, {\it{Principles of Optics}} (Cambridge University Press, Cambridge, England, 1999).


\bibitem{saleh}B. A. A. Saleh and M. C. Teich, {\it Fundamentals of Photonics} (John Wiley \& Sons, New York, 1991).

\bibitem{rayleigh}Lord Rayleigh, Philos. Mag. {\bf 8}, 261 (1879).

\bibitem{hell}S. W. Hell, Science {\bf 316}, 1153 (2007).

\bibitem{hettich}C. Hettich, Science {\bf 298}, 385 (2002).

\bibitem{Walls} P. Storey, M. Collett, and D. F. Walls, Phys. Rev. Lett. {\bf 68}, 472 (1992); Phys. Rev. A {\bf 47}, 405 (1993); R. Quadt, M. Collett, and D. F. Walls, Phys. Rev. Lett. {\bf 74}, 351 (1995).
%
\bibitem{Rempe}S. Kunze, K. Dieckmann, and G. Rempe, Phys. Rev. Lett. {\bf 78}, 2038 (1997).
%

\bibitem{Herkommer} A. M. Herkommer, W. P. Schleich, and M. S. Zubairy, J. Mod. Opt. {\bf 44}, 2507 (1997).
%

\bibitem{distance}J.-T. Chang, J. Evers, M. O. Scully, and M. S. Zubairy, Phys. Rev. A {\bf 73}, 031803(R) (2006).

\bibitem{distance2}J.-T. Chang, J. Evers and M. S. Zubairy,
Phys. Rev. A {\bf 74}, 043820 (2006).

\bibitem{gulfam}Q. Gulfam and J. Evers, J. Phys. B {\bf 43}, 045501 (2010).


\bibitem{scully}M. O. Scully and K. Dr\"uhl, Phys. Rev. A {\bf 25}, 2208 (1982).


\bibitem{complicated}S. I. Schmid and J. Evers, Phys. Rev. A {\bf} 81, 063805 (2010).

\bibitem{rabi}Z. Liao, M. Al-Amri, and M. S. Zubairy,
Phys. Rev. Lett. {\bf 105}, 183601 (2010).

\bibitem{kapale}K. T. Kapale, G. S. Agarwal,
Opt. Lett. {\bf 35}, 2792 (2010).



\bibitem{prop1} E. Paspalakis, A. F. Terzis, and P. L. Knight,  J. Mod. Opt. {\bf 52}, 1685 (2005).

\bibitem{prop2a} G. S. Agarwal and K. T. Kapale,  J. Phys. B {\bf 39}, 3437 (2006).



\bibitem{prop2b} D. D. Yavuz and N. A. Proite,  Phys. Rev. A 76, 041802 (2007).

\bibitem{prop2c} A. V. Gorshkov, L. Jiang, M. Greiner, P. Zoller, and M. D. Lukin,  Phys. Rev. Lett. 100, 093005 (2008).

\bibitem{prop2d} M. Kiffner, J. Evers, and M. S. Zubairy,  Phys. Rev. Lett. {\bf 100}, 073602 (2008).

\bibitem{prop3}  H. S. Park, S. K. Lee, and J. Y. Lee, Opt. Express {\bf 16}, 21982 (2008).

\bibitem{rabi-litho} Z. Liao, M. Al-Amri, and M. S. Zubairy, Phys. Rev. Lett. {\bf 105}, 183601 (2010)

\bibitem{prop4} T. N. Dey  and G. S. Agarwal, Phys. Rev. A {\bf 76}, 015802 (2007).

\bibitem{prop5} A. G. Truscott, M. E. J. Friese, N. R. Heckenberg, and H. Rubinsztein-Dunlop, Phys. Rev. Lett. {\bf 82}, 1438 (1999).

\bibitem{theory1} R. Kapoor and G. S. Agarwal, Phys. Rev. A {\bf 61}, 053818 (2000).

\bibitem{theory2} J. A. Andersen, M. E. J. Friese, A. G. Truscott, Z. Ficek, P. D. Drummond, N. R. Heckenberg, and H. Rubinsztein-Dunlop, Phys. Rev. A {\bf 63}, 023820 (2001).

\bibitem{prop6} P. K. Vudyasetu, D. J. Starling, and J. C. Howell, Phys. Rev. Lett. {\bf 102}, 123602 (2009).

\bibitem{prop7} H. Li, V. A. Sautenkov, M. M. Kash, A. V. Sokolov, G. R. Welch, Y. V. Rostovtsev, M. S. Zubairy, and M. O. Scully, Phys. Rev. A {\bf 78}, 013803 (2008).

\bibitem{prop8} O. Firstenberg, M. Shuker, N. Davidson, and A. Ron, Phys. Rev. Lett. {\bf 102}, 043601 (2009).

\bibitem{prop9} O. Firstenberg, P. London, M. Shuker, A. Ron, and N. Davidson, Nature Phys.  {\bf 5}, 665 (2009).



\bibitem{wg1} F. Benabid, J. C. Knight, G. Antonopoulos, and P. S. J. Russell, Science {\bf 298}, 399 (2002).

\bibitem{wg2}S. Ghosh, A. R. Bhagwat, C. K. Renshaw, S. Goh, A. L. Gaeta, and B. J. Kirby, Phys. Rev. Lett. {\bf 97}, 023603 (2006).

\bibitem{focusing1} R. R. Moseley, S. Shepherd, D. J. Fulton, B. D. Sinclair, and M. H. Dunn, Phys. Rev. A {\bf 53}, 408 (1996).

\bibitem{focusing2} R. R. Moseley, S. Shepherd, D. J. Fulton, B. D. Sinclair, and M. H. Dunn, Phys. Rev. Lett. {\bf 74}, 670 (1995).

\bibitem{focusing3} D. R. Walker, D. D. Yavuz, M. Y. Shverdin, G. Y. Yin, A. V. Sokolov, and S. E. Harris, Opt. Lett. {\bf 27}, 2094.

\bibitem{focusing4} N. A. Proite, B. E. Unks, J. T. Green, and D. D. Yavuz, Phys. Rev. A {\bf 77}, 023819 (2008).

\bibitem{dey1}T. N. Dey, G. S. Agarwal, \ol {\bf 34}, 3199 (2009)

\bibitem{kash}M. M. Kash, V. A. Sautenkov, A. S. Zibrov, L. Hollberg, G. R. Welch, M. D. Lukin,Y. Rostovtsev,E. S Fry and M. O. Scully, \prl {\bf 82}, 5229 (1999).
\bibitem{kochar}O. Kocharovskaya, Y. Rostovtsev, and M. O. Scully, \prl {\bf 86} 628 (2001).

\bibitem{Shen}A. D. Bandrauk, H. Shen, Journal of Physics A: Mathematical and General  {\bf 27}, 7747 (1994).


\end{thebibliography}
\end{document}